\documentclass[fleqn,usenatbib]{mnras}

\usepackage{newtxtext,newtxmath}

\usepackage[T1]{fontenc}
\usepackage{ae,aecompl}

\usepackage{graphicx}	
\usepackage{amsmath}	
\usepackage{epsfig}
\usepackage{epstopdf}

\def\be{\begin{equation}}
\def\ee{\end{equation}}

\usepackage{graphicx}
\usepackage{dcolumn}
\usepackage{bm}
\usepackage{xcolor}
\usepackage{hyperref}
\usepackage{mathtools}
\usepackage{physics}
\usepackage{booktabs}
\usepackage{siunitx}
\usepackage{graphicx}
\usepackage{float}
\usepackage{tabularx}

\hypersetup{
	colorlinks=true,
	linkcolor=blue,
	citecolor=blue,
	urlcolor=blue
}

\newcommand{\figplaceholder}[2]{%
	\IfFileExists{#1}{%
		\includegraphics[width=\linewidth]{#1}%
	}{%
		\fbox{%
			\parbox[c][4.0cm][c]{0.92\linewidth}{%
				\centering
				\textbf{Figure placeholder}\\[0.5em]
				#2\\[0.5em]
				\texttt{#1}
			}%
		}%
	}%
}

\title[Initial conditions \& Early Bright Galaxies]{Localized Enhancements of the Primordial Power Spectrum and the Abundance of Early Bright Galaxies}
\author[Shaghayegh Yavari, Hamed Kameli, Shant Baghram]{Shaghayegh Yavari $^{1}$
Hamed Kameli $^{2}$
~and Shant Baghram $^{1,3}$ \thanks{baghram@sharif.edu}
\\
$^{1}$
Department of Physics, Sharif University of
Technology, P.~O.~Box 11155-9161, Tehran, Iran\\\
$^2$Department of Physics, Center for Exact and Natural Sciences, Federal University of Paraíba, 58059-970, João Pessoa, Brazil \\
$^3$Research Center for High Energy Physics, Department of Physics, Sharif University of Technology, Tehran 11155-9161, Iran
}
\date{Accepted XXX. Received YYY; in original form ZZZ}
\pubyear{2026}
\begin{document}
\label{firstpage}
\pagerange{\pageref{firstpage}--\pageref{lastpage}}
\maketitle
\begin{abstract}
Recent observations of bright galaxies at high redshifts by the James Webb Space Telescope have motivated renewed interest in cosmological mechanisms that can enhance the abundance of early massive dark matter halos. We study a phenomenological modification of the primordial curvature power spectrum in the form of a localized Gaussian enhancement. The modified spectrum is propagated to the linear matter power spectrum, the smoothed density variance, and the Sheth--Tormen halo mass function. We show that a bump centered at small scales can selectively enhance the abundance of rare high-redshift halos and, through a fixed semi-analytic mapping between halo mass and ultraviolet luminosity, shift the bright end of the UV luminosity function toward JWST measurements. The enhancement becomes more pronounced at higher redshift due to the exponential sensitivity of the rare-halo tail to the variance. We emphasize that the effect is limited by small-scale power-spectrum constraints, particularly Lyman-\(\alpha\) forest data, and should be interpreted as a partial cosmological contribution rather than a complete solution to the JWST galaxy tension.
\end{abstract}




\section{Introduction}
\label{sec:introduction}

The James Webb Space Telescope (JWST) has substantially extended the
observational frontier of galaxy formation into the first few hundred
million years of cosmic history. Deep Near-InfraRed Camera (NIRCam) imaging has identified
large samples of candidate galaxies at \(z\gtrsim 10\), while Near-Infrared Spectrograph (NIRSpec)
follow-up has now confirmed luminous systems beyond \(z\simeq 14\)
\citep{2024Natur.633..318C,2024ApJ...969L...2F,2025ApJ...980..138H}.
A particularly striking result is the relatively high abundance of the bright end of the rest-frame ultraviolet
luminosity function (UVLF). Galaxies with
\(M_{\rm UV}\lesssim-20\) appear substantially more common at
\(z\gtrsim 10\) than predicted by many galaxy-formation models developed
before JWST, with some measurements reporting an excess of one or more
orders of magnitude \citep{2024MNRAS.531.2615C}. Although the earliest photometric samples were
affected by redshift uncertainties and low-redshift interlopers,
spectroscopic surveys continue to find luminous galaxies at very early
epochs. The emerging tension should therefore be understood primarily
as a challenge to our understanding of early galaxy formation or our cosmological model \citep{2023NatAs...7..731B}.

There are two broad ways of interpreting this excess. The first is to
retain the standard cosmological halo population and modify the
astrophysics connecting dark-matter (DM) halos to their observable
luminosities. One possibility is that star formation was substantially
more efficient at cosmic dawn than at later times \citep{2023A&A...677L...4P}. The very high gas
densities and short dynamical times expected in early massive halos may
allow rapid, feedback-free starbursts in which a large fraction of the
available gas is converted into stars before stellar winds and
supernovae can regulate the process \citep{2023MNRAS.523.3201D}. 

Other astrophysical ingredients can modify the amount of UV light
produced for a given stellar or halo mass. A top-heavy stellar initial
mass function increases the abundance of short-lived massive stars and
therefore the UV luminosity per unit stellar mass \citep{2025ApJ...980...10J}. Active Galactic Nuclei (AGN) population may differ from standard picture of galaxy evolution and affect the results \citep{2024JCAP...08..025H}.

A complementary possibility is that the abundance of the halos
themselves differs from the standard expectation. This has motivated a
range of cosmological explanations for the JWST observations.
Dynamic dark energy models can alter the expansion and linear-growth
histories and thereby change the abundance of rare halos at early
times \citep{2023JCAP...10..072A}.
Early dark energy models have likewise been explored as a mechanism for
increasing the abundance of high-redshift galaxies
\citep{2024MNRAS.533.3923S,2024EPJP..139..711W}. These approaches modify the subsequent
growth of the density field. A different possibility, and the one
considered here is to modify the spectrum from which the structures
grow in the first place \citep{2023ApJ...953L...4P,2024MNRAS.527.1381T,2025PhRvD.112l3503F}.

The abundance of rare high-redshift halos is particularly sensitive to
the amplitude of the primordial fluctuations on the corresponding
comoving scales. Several studies have therefore investigated whether
additional small-scale primordial power can ease the apparent JWST
tension. A blue-tilted primordial spectrum can accelerate early
structure formation \citep{2023MNRAS.526L..63P}.

A localized enhancement also has a broader theoretical motivation.
The nearly power-law primordial curvature spectrum of the minimal
cosmological model is characteristic of simple slow-roll inflation,
but it is not a generic prediction of all inflationary dynamics.
Transient violations of slow roll, localized structure in the inflaton
potential, changes the propagation of curvature perturbations, or
bursts of particle production during inflation can generate features
over a restricted range of wavenumbers \citep{2011PhRvD..83j3501C,2018JCAP...06..004P}.

We therefore use a
Gaussian enhancement as a phenomenological description of localized
excess primordial power. The Gaussian form is not intended to identify
a unique inflationary mechanism; rather, it provides a controlled way
to vary the amplitude, width, and characteristic scale of the feature
while leaving scales sufficiently far from it nearly unchanged.

This type of modification has also been considered previously in the context of small-scale structure. \cite{2022MNRAS.511.1601K} studied a Gaussian excess in the initial curvature spectrum. They
showed that the modification can increase the abundance of halos near
the corresponding galactic mass scale while reducing the population of
lower-mass subhalos, providing a possible route to alleviating the
Too-Big-to-Fail and missing-satellite problems. Subsequent work extended
this picture to conditional halo statistics, mergers, and structure
formation in underdense environments  \citep{2023MNRAS.526.1495P,2025MNRAS.542.3331K}.

In this work, we apply this scale-selective modification of the initial
conditions (IC)s to the population of galaxies observed at cosmic dawn.
We ask whether a localized primordial feature can
increase the abundance of the dark-matter halos capable of hosting
UV-bright galaxies. We introduce a Gaussian feature in the primordial
curvature spectrum and propagate it to the linear matter power
spectrum, mass variance, and halo mass function. We then determine how
the response depends on the position of the feature and on redshift,
constrain the allowed modification using measurements of the matter
power spectrum, and map the resulting halo population to the observed
UVLF. We consider the small scale constraints from Lyman-$\alpha$ data to introduce the wavenumber of the enhancement. We fit the UVLF with the modified halo mass function with the corresponding halo to galaxy relation. We also show the modified halo number density in the context of non-Markov Excursion set theory. Showing that this enhancement is in the right direction to solve the observed tension. The structure of this work is as below: In Section \ref{sec:framework} we discuss the theoretical background explaining the modified halo abundance. In Section \ref{sec:halo_response}, we show the effect of the bumpy power spectrum . In Section \ref{sec:assembly_history}, we study the effect of the modified IC on progenitor distributions and halo assembly. In Section \ref{sec:constraints}, we compare the model with observation UVLF. In Section \ref{sec:uvlf_model}, we show the relation of halo mass function and UVLF. Finally, in Section \ref{sec:conclusions}, we conclude and have our discussion for future prospects.

\section{Dark Matter Halo Abundance from Modified Initial Conditions}
\label{sec:framework}
In this section, first we discuss the procedure of finding the halo abundance. In the second subsection, we show the effect of the localized Gaussian feature on power spectrum. 

\subsection{Halo-Abundance Framework}
\label{subsec:est_hmf}
The DM halo mass function provides the first step in connecting the linear matter power spectrum to the observable population of galaxies residing in DM halos. It describes the comoving number density of DM halos as a function of halo mass and redshift.  Modifications to the initial density fluctuations can affect galaxy abundances through their impact on the halo population. To describe the formation and abundance of halos, we first adopt the EST framework, which relates the statistics of the initial density fluctuations to the collapse of matter into halos \citep{1991ApJ...379..440B}. In this framework, the dependence on halo mass is introduced through the smoothing scale of the density field, while the time or redshift dependence is encoded in the evolution of the critical density threshold for collapse
\citep{2007IJMPD..16..763Z,2017PhRvD..96d3524N}.

Within this framework, one of the key quantities is the
variance of the linear density field smoothed on the scale associated
with a halo of mass \(M\). For a spherical real-space top-hat filter,
the smoothing radius $R$ and halo mass are related by
\(M=(4\pi/3)\bar{\rho}_{m,0}R^3\), where
\(\bar{\rho}_{m,0}\) is the present-day mean comoving matter density.
The variance is

\begin{equation}
S\equiv	\sigma^2(M)
	=
	\frac{1}{2\pi^2}
	\int_0^\infty
	k^2 P_m(k,0)
	W_{\rm TH}^2(kR)\,dk ,
	\label{eq:mass_variance}
\end{equation}

where
\(W_{\rm TH}(x)=3(\sin x-x\cos x)/x^3\) is the Fourier of top-hat filter.
Equation~\eqref{eq:mass_variance}, provides the direct link between a change in the matter power spectrum and the mass scales on
which halo abundances are modified.

We evaluate the variance using the linear density field extrapolated
to \(z=0\). The redshift dependence therefore enters through the
collapse threshold
\(\delta_c(z)=\delta_{c,0}/D(z)\), where $D(z)$ is the growth function normalized to unity in present time with
\(\delta_{c,0}\simeq1.686\) \citep{1972ApJ...176....1G}.
The resulting variance and collapse threshold determine the
mass and redshift dependence of halo abundances. To describe this
abundance, one should calculate the statistics of first upcrossing in EST models \citep{2018MNRAS.478.5296N}. In this work we neglect the complications of first upcrossing issue and we adopt the Sheth-Tormen halo mass function, which is well calibrated to fit cosmological simulations
\citep{1999MNRAS.308..119S}:

\begin{equation}
	\frac{dn(M,z)}{d\ln M}
	=
	\frac{\bar{\rho}_{m,0}}{M}
	f_{\rm ST}(\nu)
	\left|
	\frac{d\ln\sigma^{-1}}{d\ln M}
	\right|,
	\label{eq:general_hmf}
\end{equation}

where the peak height is
\(\nu(M,z)=\delta_c(z)/\sigma(M)\). The corresponding multiplicity
function is

\begin{equation}
	f_{\rm ST}(\nu)
	=
	A
	\sqrt{\frac{2a}{\pi}}
	\left[
	1+(a\nu^2)^{-p}
	\right]
	\nu
	\exp\left(-\frac{a\nu^2}{2}\right),
	\label{eq:sheth_tormen}
\end{equation}

with \(A=0.3222\), \(a=0.707\), and \(p=0.3\).
Unless otherwise stated, this prescription is used for all halo
abundance and UVLF calculations. Note that  cosmological models affect the matter power spectrum, through the initial conditions and the evolution of the dark sector through the growth function \citep{2025MNRAS.542.3331K,2026MNRAS.546ag269C}. This means that the abundance of DM halos is an indirect probe of cosmological models.

\subsection{Localized Gaussian Feature in the Primordial Spectrum}
\label{subsec:gaussian_feature}

We introduce a Gaussian enhancement in the primordial curvature power spectrum in order to increase the initial fluctuations over a restricted range of comoving wave-numbers while leaving scales far from the feature approximately unchanged. The enhancement is confined to small scales, where observational constraints are less restrictive. The power spectrum in larger scales is kept unchanged to remain consistent with the strong observational constraints on these scales \citep{2012JPhCS.375c2012S,2020MNRAS.494.4907K}. The modified spectrum with Gaussian bump $\mathcal{P}_{\mathcal R}^{\rm G}$ is
\begin{equation}
	\mathcal{P}_{\mathcal R}^{\rm G}(k)
	=
	\mathcal{P}_{\mathcal R}^{\rm std}(k)
	\mathcal{G}(k),
	\label{eq:modified_primordial_spectrum}
\end{equation}

where the standard spectrum is defined as
\(\mathcal{P}_{\mathcal R}^{\rm std}(k)
=A_s(k/k_p)^{n_s-1}\), and

\begin{equation}
	\mathcal{G}(k)
	=
	1+
	\frac{A_b}{\sqrt{2\pi}\sigma_b}
	\exp\left[
	-\frac{(k-k_\ast)^2}{2\sigma_b^2}
	\right],
	\label{eq:gaussian_feature_factor}
\end{equation}
where \(k_\ast\) sets the location of the feature,
\(\sigma_b\) determines its width, and \(A_b\) controls its amplitude.

The primordial curvature spectrum is transferred to the linear matter
power spectrum through the transfer function and growth factor. In
linear theory matter power spectrum is related to the primordial power $\mathcal{P}_{\mathcal R}(k)$ via
\begin{equation}
	P_m(k,z)= \frac{8\pi^2}{25}\frac{k}{\Omega^2_{m,0}H_{0}^{4}} \,T^2(k)D^2(z)\mathcal{P}_{\mathcal R}(k),
\end{equation}
with the same growth and transfer functions in
Gaussian-feature cosmologies as in standard model. We evaluate \(T(k)\) using the
Eisenstein--Hu transfer function \citep{1998ApJ...496..605E}.
Because the two models otherwise share the same background cosmology,
transfer function, and growth history, we can see that
\(
{P_m^{\rm G}(k,z)}/
{P_m^{\rm std}(k,z)}
=
\mathcal{G}(k).\)
The primordial feature therefore appears as a localized enhancement
of the linear matter power spectrum. Through
equation ~\eqref{eq:mass_variance}, this changes both the amplitude and the
mass dependence of \(\sigma(M)\), and hence the halo abundance. We now
examine how this response depends on the location of the feature and
on redshift.

\section{Effect of the Gaussian Feature on Halo Abundance}
\label{sec:halo_response}

\subsection{Redshift Amplification of the Halo Response}
\label{subsec:redshift_amplification}

The same change in the variance produces a much larger fractional
change in halo abundance at high redshift. This behavior follows from
the exponential dependence of the Sheth--Tormen multiplicity function
on the peak height. At fixed halo mass, the ratio of the two mass
functions can be written directly from equation.~\eqref{eq:general_hmf} as

\begin{equation}
	\mathcal{R}_{\rm HMF}
	=
	\frac{
		\left|
		d\ln\sigma_{\rm G}^{-1}/d\ln M_h
		\right|
	}{
		\left|
		d\ln\sigma_{\rm std}^{-1}/d\ln M_h
		\right|
	}
	\,
	\frac{
		f_{\rm ST}(\nu_{\rm G})
	}{
		f_{\rm ST}(\nu_{\rm std})
	},
	\label{eq:st_ratio_compact}
\end{equation}
where subscript of variance and height parameter refers to Gaussian bump and standard model respectively.
Using equation.~\eqref{eq:sheth_tormen}, this becomes

\begin{align}
	\mathcal{R}_{\rm HMF}(M_h,z)
	&=
	\frac{
		\left|
		d\ln\sigma_{\rm G}^{-1}/d\ln M_h
		\right|
	}{
		\left|
		d\ln\sigma_{\rm std}^{-1}/d\ln M_h
		\right|
	}
	\left(
	\frac{S_{\rm std}}{S_{\rm G}}
	\right)^{1/2}
	\nonumber\\
	&\quad\times
	\frac{
		1+\left[a\delta_c^2(z)/S_{\rm G}\right]^{-p}
	}{
		1+\left[a\delta_c^2(z)/S_{\rm std}\right]^{-p}
	}
	\nonumber\\
	&\quad\times
	\exp\left\{
	\frac{a\delta_c^2(z)}{2}
	\left[
	\frac{1}{S_{\rm std}}
	-
	\frac{1}{S_{\rm G}}
	\right]
	\right\}.
	\label{eq:exact_st_hmf_ratio}
\end{align}

At fixed mass, \(S_{\rm std}\) and \(S_{\rm G}\) are independent of
redshift in our convention, while
\(\delta_c(z)=\delta_{c,0}/D(z)\) increases toward earlier times.
The first factors in equation.~\eqref{eq:exact_st_hmf_ratio} therefore vary
much more weakly with redshift than the exponential term. In a
high-\(\nu\) regime,

\begin{equation}
	\ln\mathcal{R}_{\rm HMF}(M_h,z)
	\simeq
	\frac{a\delta_c^2(z)}{2}
	\left[
	\frac{1}{S_{\rm std}(M_h)}
	-
	\frac{1}{S_{\rm G}(M_h)}
	\right].
	\label{eq:high_nu_log_ratio}
\end{equation}

For masses at which the feature gives
\(S_{\rm G}>S_{\rm std}\), the coefficient multiplying
\(\delta_c^2(z)\) is positive. The enhancement of rare halos is
therefore amplified rapidly toward high redshift. This explains why
the same localized change in the initial power spectrum produces a
moderate HMF response at lower redshift but a much larger fractional
change once the relevant halos lie deep in the rare-event tail.

\begin{figure*}
	\centering
	\includegraphics[width=\textwidth]
	{\detokenize{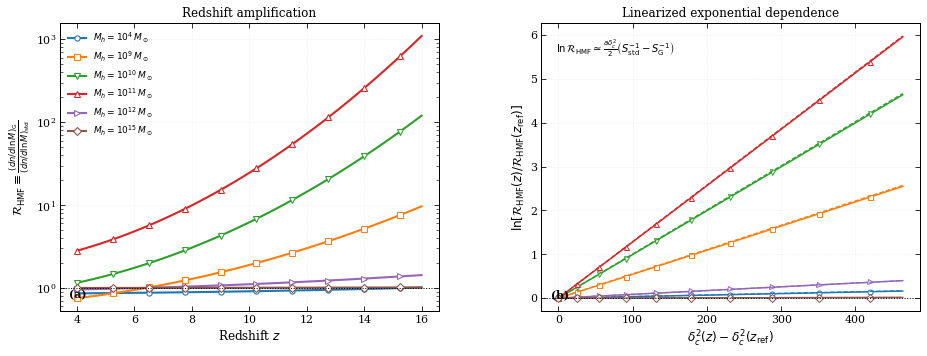}}
	\caption{
		Redshift amplification of the Gaussian-to-standard HMF ratio for
		\(k_\ast=4.15\,h \mathrm{Mpc}^{-1}\),
		\(A_b=8\,h \mathrm{Mpc}^{-1}\), and \(\sigma_b=0.4\,h\mathrm{Mpc}^{-1}\).
		Left: full Sheth--Tormen ratios at fixed halo masses from
		\(10^4\) to \(10^{15}\,M_\odot\), together with the dominant
		high-\(\nu\) exponential prediction normalized at the reference
		redshift $z=4$. Right: logarithmic HMF ratio as a function of the change
		in \(\delta_c^2\) relative to the reference redshift. The agreement
		shows that the increasingly strong response toward high redshift
		is driven primarily by the exponential rare-halo tail.
	}
	\label{fig:exponential_growth}
\end{figure*}

Fig.~\ref{fig:exponential_growth} demonstrates this interpretation. The
left panel shows HMF ratio with the dominant
high-\(\nu\) exponential contribution at several fixed halo masses.
The right panel isolates the predicted dependence on
\(\delta_c^2(z)\). The approximately linear behavior confirms that
the rapid redshift evolution of the ratio is controlled primarily by
the exponential tail of the Sheth--Tormen mass function.

\begin{figure}
	\centering
	\includegraphics[width=0.92\linewidth]
	{\detokenize{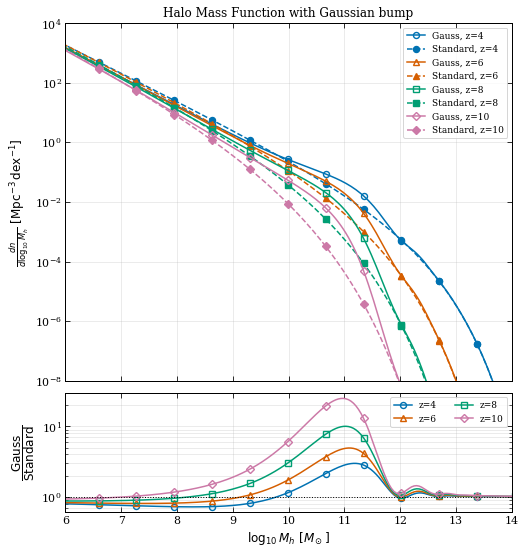}}
	\caption{The upper panel:
		Absolute differential halo mass functions for the standard
		spectrum (dashed curves) and Gaussian-feature spectrum
		(solid curves) at \(z=4,\ 6,\ 8,\) and \(10\).
		The Gaussian feature is centered at
		\(k_\ast=4.15\,\mathrm{hMpc}^{-1}\), with
		\(A_b=8 \,\mathrm{hMpc}^{-1}\) and \(\sigma_b=0.4\,\mathrm{hMpc}^{-1}\).
		The difference between the two models becomes increasingly
		pronounced around the bump and at higher redshift. The lower panel: the ratio of the HMF in two models.
	}
	\label{fig:absolute_hmf_redshifts}
\end{figure}

Large fractional enhancements must nevertheless be interpreted
together with the absolute abundance. A large value of
\(\mathcal{R}_{\rm HMF}\) can occur in a region where both models
predict very few halos. Fig.~\ref{fig:absolute_hmf_redshifts}
therefore shows the absolute differential HMF for the standard and
Gaussian-feature spectra.

The effect is particularly
pronounced at high redshift, where massive halos correspond to larger
peak heights. Thus, the localized variance enhancement does not
produce a uniform increase in halo abundance, its impact; localized around the bump scale, is stronger
for rare, massive objects and grows rapidly toward earlier epochs.

\subsection{Scale Dependence of the Halo Response}
\label{subsec:kstar_dependence}

Decreasing \(k_\ast\) shifts the halo response toward larger masses,
as expected from the approximate scaling
\(M_{\rm feature}\propto k_\ast^{-3}\). Beyond this shift, the
different feature locations also produce different strengths of the
exponential response. The relevant variance combination appearing in equation~\eqref{eq:high_nu_log_ratio} is

\begin{equation}
	\mathcal{D}(M_h,k_\ast)
	\equiv
	\frac{1}{S_{\rm std}(M_h)}
	-
	\frac{1}{S_{\rm G}(M_h,k_\ast)}
	\label{eq:variance_driver}
\end{equation}

\begin{figure}
	\centering
	\includegraphics[width=\linewidth]
	{\detokenize{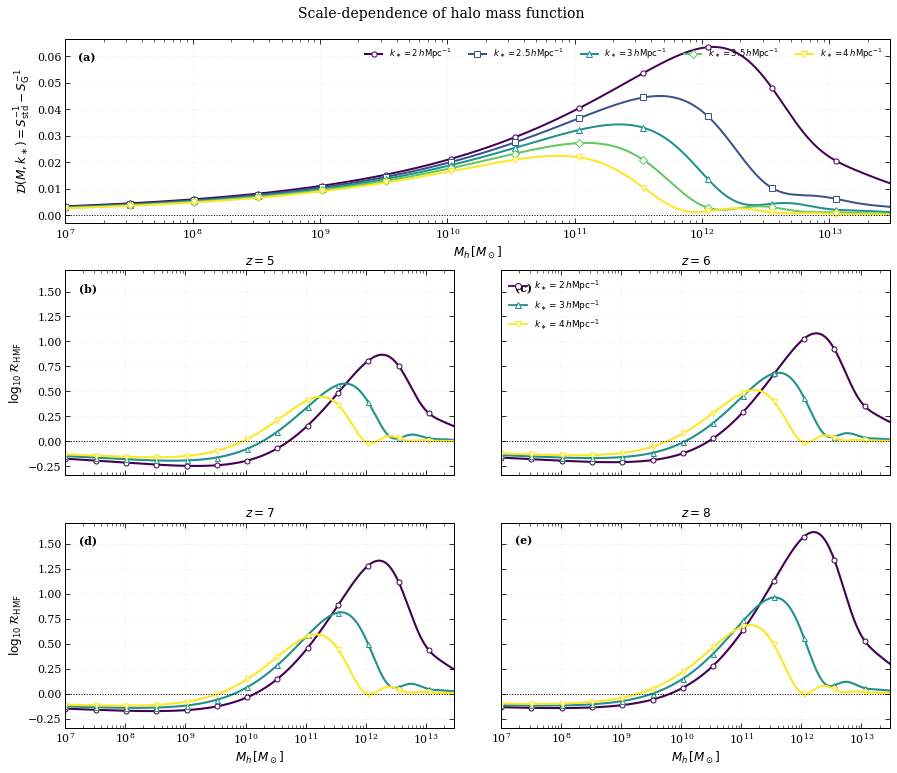}}
	\caption{
		Scale dependence of the Gaussian-feature halo response.
		Panel~(a) shows the variance combination
		\(\mathcal{D}(M_h,k_\ast)
		=S_{\rm std}^{-1}-S_{\rm G}^{-1}\)
		for \(k_\ast=2,\ 2.5,\ 3,\ 3.5,\) and
		\(4\,h \mathrm{Mpc}^{-1}\).
		Panels~(b)--(e) show
		\(\log_{10}\mathcal{R}_{\rm HMF}\) at
		\(z=5,\ 6,\ 7,\) and \(8\) for
		\(k_\ast=2,\ 3,\) and \(4\,h \mathrm{Mpc}^{-1}\).
		Solid curves show the full Sheth--Tormen HMF ratio, while dashed
		curves show only the exponential contribution
		\(\exp\!\left[
		a\delta_c^2(z)
		(S_{\rm std}^{-1}-S_{\rm G}^{-1})/2
		\right]\).
		The feature parameters are fixed to
		\(A_b=8\ \,h \mathrm{Mpc}^{-1}\) and
		\(\sigma_b=0.4\,h\mathrm{Mpc}^{-1}\).
	}
	\label{fig:hmf_kstar_scan}
\end{figure}

Panel~(a) of Fig.~\ref{fig:hmf_kstar_scan} shows
\(\mathcal{D}(M_h,k_\ast)\) for several feature locations. For the
fixed amplitude and width adopted here, decreasing \(k_\ast\) produces
a progressively larger maximum value of \(\mathcal{D}\), while moving
that maximum toward higher halo mass. Since \(\mathcal{D}\) is the
coefficient of the redshift-dependent exponential contribution in
equation~\eqref{eq:high_nu_log_ratio}, the lower-\(k_\ast\) cases consequently
develop the stronger HMF enhancement seen in panels~(b)--(e).

The dashed curves in panels~(b)--(e) isolate the exponential factor,
whereas the solid curves show the complete Sheth--Tormen ratio of
Eq.~\eqref{eq:exact_st_hmf_ratio}. The exponential term reproduces the
broad location and growth of the enhancement, but not its full
mass-dependent structure. The sharper peaks, the suppression below
unity on the low-mass side, and the smaller secondary features in the
full result arise from the remaining factors in
Eq.~\eqref{eq:exact_st_hmf_ratio}, including the change in
\(\left|d\ln\sigma^{-1}/d\ln M_h\right|\).

\section{Progenitor Distributions, Halo Assembly and Bias}
\label{sec:assembly_history}
In this section, we study two pillars of halo statistics. In the first subsection we study the progenitor distribution and in the second subsection we address the bias parameter.
\subsection{Progenitor distribution}

The HMF describes the abundance of halos at a given
epoch, but does not specify how those halos were assembled. We therefore
use the conditional excursion-set distribution to examine the
progenitors of a halo of mass \(M_0\) at redshift \(z_0\) \citep{1993MNRAS.262..627L,2022MNRAS.511.1601K}. For a
progenitor of mass \(M_1<M_0\) at \(z_1>z_0\), we define $\Delta S=S(M_1)-S(M_0)$ and $\Delta\delta=\delta_c(z_1)-\delta_c(z_0)$.
Within the spherical extended Press--Schechter approximation \citep{2010gfe..book.....M}, the
conditional first-crossing distribution is

\begin{equation}
	f(\Delta S,\Delta\delta)
	=
	\frac{\Delta\delta}
	{\sqrt{2\pi}(\Delta S)^{3/2}}
	\exp\left[
	-\frac{\Delta\delta^2}{2\Delta S}
	\right],
	\label{eq:conditional_eps}
\end{equation}

which gives the progenitor distribution per logarithmic mass interval

\begin{equation}
	\frac{dN_{\rm prog}}{d\log_{10}M_1}
	=
	\frac{M_0}{M_1}
	f(\Delta S,\Delta\delta)
	\left|
	\frac{dS}{d\log_{10}M_1}
	\right|.
	\label{eq:conditional_progenitor}
\end{equation}

Here \(dN_{\rm prog}/d\log_{10}M_1\) is the expected number of
progenitors per descendant halo.
To show which progenitor masses actually dominate the ancestral mass
budget, we also consider \(
\frac{M_1}{M_0}
\frac{dN_{\rm prog}}{d\log_{10}M_1}\).

\begin{figure}
	\centering
	\includegraphics[width=\linewidth]
	{\detokenize{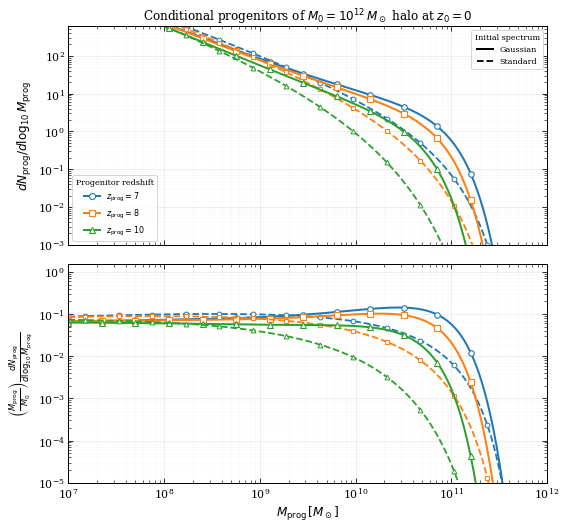}}
	\caption{
		Conditional progenitor distributions for a
		\(M_0=10^{12}\,M_\odot\) halo at \(z_0=0\) (same Gaussian parameters as Fig. \ref{fig:absolute_hmf_redshifts}~).
		Solid and dashed curves denote the Gaussian-feature and standard
		models, respectively. \\ Top panel: expected progenitor number per
		logarithmic mass interval at \(z_{\rm prog}=7,\ 8,\) and \(10\).
		Bottom panel: the corresponding contribution to the descendant mass
		fraction.
	}
	\label{fig:prog1}
\end{figure}

Fig.~\ref{fig:prog1} shows these quantities for a
\(10^{12}\,M_\odot\) descendant at \(z_0=0\) as an example. Relative to the standard
model, the Gaussian feature suppresses the low-mass part of the
conditional progenitor distribution while enhancing its high-mass tail as the redshift of the progenitor under study increases. The lower panel shows that this is also a redistribution of the
ancestral mass. At the same progenitor redshift, a larger fraction of
the final halo mass is already contained in relatively massive
progenitors in the Gaussian model. The contrast becomes stronger
toward earlier epochs. The feature therefore produces a less
fragmented and more advanced progenitor population around the mass scale corresponding to the Gaussian bump scale.

A complementary measure is the half-mass formation redshift \citep{2010gfe..book.....M}. Since
there can be at most one progenitor with \(M_1>M_0/2\) at each redshift, the integral

\begin{equation}
	P_{1/2}(z|M_0,z_0)
	=
	\int_{M_0/2}^{M_0}
	\frac{dN_{\rm prog}}{dM_1}\,dM_1
	\label{eq:half_mass_probability}
\end{equation}

is the probability that the main progenitor has already exceeded half
of the descendant mass. We define the median formation redshift by

\begin{equation}
	P_{1/2}(z_{\rm half}|M_0,z_0)=\frac{1}{2},
\end{equation}

and compare the two models through

\begin{equation}
	\Delta z_{\rm half}
	=
	z_{\rm half}^{\rm G}-z_{\rm half}^{\rm std}.
	\label{eq:delta_z50}
\end{equation}

\begin{figure}
	\centering
	\includegraphics[width=\linewidth]
	{\detokenize{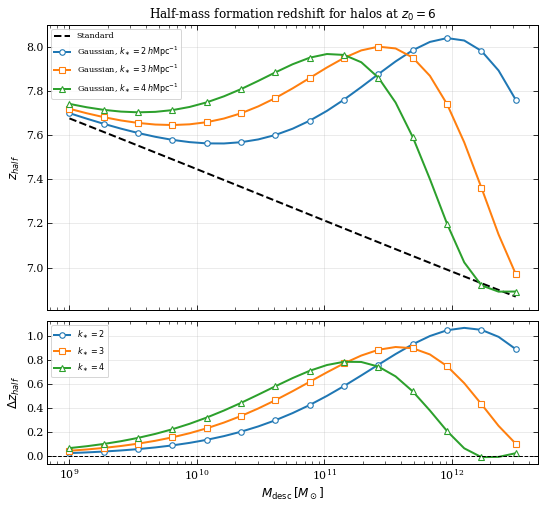}}
	\caption{
		Median half-mass formation redshift for descendants at \(z_0=6\).
		Top: \(z_{\rm half}\) for the standard model and Gaussian features with
		\(k_\ast=2,\ 3,\) and \(4\,h \mathrm{Mpc}^{-1}\).
		Bottom: the corresponding shift
		\(\Delta z_{\rm half}=z_{\rm half}^{\rm G}-z_{\rm half}^{\rm std}\).
		Positive values indicate earlier half-mass assembly in the
		Gaussian model. Increasing \(k_\ast\) moves the strongest assembly
		response toward smaller descendant masses.
	}
	\label{fig:prog2}
\end{figure}

Fig.~\ref{fig:prog2} shows that the standard model retains the usual
hierarchical trend in which more massive descendants assemble later.
The Gaussian feature instead produces a localized advance in the
half-mass formation epoch. 

Together, the two diagnostics show that the localized primordial
feature changes not only halo abundance but also the route by which
halos assemble. Ancestral mass is transferred toward more massive
progenitors according to the scale of the Gaussian bump, causing the main progenitor to reach
half of the final halo mass sooner over a characteristic descendant
mass range. 

\subsection{Halo Bias and Spatial Clustering: Breaking the Degeneracy with Astrophysics}
\label{subsec:bias_clustering}

A fundamental challenge in interpreting the overabundance of bright, high-redshift galaxies observed by JWST is the inherent degeneracy between cosmological solutions and non-standard astrophysical mechanisms. For instance, increasing the star formation efficiency $f_*$ or invoking a top-heavy initial mass function (IMF) shifts the UV luminosity function to brighter magnitudes by allowing lower-mass, more abundant halos to host luminous galaxies. Conversely, the primordial power enhancement considered here increases the baseline abundance of massive host halos at fixed standard astrophysical conversion efficiency. 

Fortunately, spatial clustering and linear halo bias provide an independent observational probe capable of breaking this degeneracy. Within the EST framework for ellipsoidal collapse, the deterministic linear halo bias $b(M, z)$ is expressed in terms of the peak height $\nu(M, z) \equiv \delta_c(z)/\sqrt{S(M)}$ via the Sheth--Tormen relation \citep{1999MNRAS.308..119S, 2018PhR...733....1D}:
\begin{equation}
	b(M, z) = 1 + \frac{a\,\nu^2 - 1}{\delta_c(z)} + \frac{2\,p/\delta_c(z)}{1 + (a\,\nu^2)^p},
	\label{eq:halo_bias}
\end{equation}
where $a = 0.707$, $p = 0.3$, and $\delta_c(z) = 1.686 / D(z)$ is the linear collapse threshold.

In our model, the introduction of the Gaussian feature modifies the smoothed mass variance, $S_G(M) > S_{\mathrm{std}}(M)$, on scales resonant with $k_*$. At fixed halo mass $M$ and redshift $z$, this enhancement increases the variance, thereby \textit{reducing} the effective peak rarity:
\begin{equation}
	\nu_G(M, z) = \frac{\delta_c(z)}{\sqrt{S_G(M)}} < \nu_{\mathrm{std}}(M, z).
	\label{eq:nu_shift}
\end{equation}
Because the halo bias is a monotonically increasing function of peak rarity in the high-$\nu$ regime ($\nu \gg 1$), the Gaussian bump leads to a distinctive prediction: The host halos become less extreme statistical fluctuations of the density field, resulting in a slightly lower intrinsic halo bias compared to standard $\Lambda\mathrm{CDM}$, $b_G(M, z) < b_{\mathrm{std}}(M, z)$.
	At fixed UV luminosity $M_{\mathrm{UV}}$, in an astrophysical scenario where bright galaxies are explained by elevated star formation efficiency ($f_* \gg 0.1$), these galaxies populate low-mass halos ($M \sim 10^9\text{--}10^{10}\,h^{-1}M_\odot$) which have substantially lower bias ($b \sim 4\text{--}6$ at $z \sim 10$). In contrast, in our modified initial condition framework, the luminous galaxies continue to reside in genuinely massive potential wells ($M \sim 10^{10.5}\text{--}10^{11.5}\,h^{-1}M_\odot$), which retain a strong, highly clustered bias ($b \sim 9\text{--}14$).

Consequently, the projected two-point angular correlation function $w(\theta)$ of bright $z \gtrsim 10$ galaxies offers a smoking-gun test. If the JWST bright-end excess is driven purely by stochastic or extreme starbursts in dwarf halos, the observed clustering amplitude will be comparatively weak. If, however, a localized primordial feature enhances the underlying abundance of massive structures, the bright galaxies will exhibit pronounced spatial clustering. Future wide-area spectroscopic and photometric surveys—such as the Roman Space Telescope High Latitude Survey and deep JWST Cycle 3/4 mosaic fields—will yield sufficiently large galaxy samples to measure $w(\theta)$ and angular clustering at $z > 10$, offering an orthogonal empirical test to validate or rule out a primordial origin for the JWST tension.

\section{Observational Constraints on the Primordial Feature}
\label{sec:constraints}

The parameters of the Gaussian feature cannot be varied independently
of existing measurements of the matter power spectrum. Different observational probes constrain different ranges of wavenumber. The
largest scales are primarily restricted by the CMB, while galaxy clustering, weak lensing, and especially the
Ly$\alpha$ forest extend the comparison toward smaller physical
scales \citep{2020A&A...641A...6P}. We use the compilation of the linear three-dimensional matter
power spectrum presented by Chabanier et al.~\cite{2019MNRAS.489.2247C},
which combines these complementary measurements at a common reference
redshift.

The location of the feature is particularly important for the present
application. Moving the bump toward smaller \(k_\ast\) shifts its
influence to larger spatial scales and, consequently, toward larger
halo masses. Such a shift would be favorable for enhancing the hosts of
bright high-redshift galaxies. At the same time, however, it increases
the overlap of the modified spectrum with the scales constrained by the
Ly$\alpha$-forest measurements \citep{2009MNRAS.399L..39V}. The feature therefore cannot be moved
indefinitely toward smaller wavenumbers without producing an increasingly
visible excess of small-scale matter power.

\begin{figure}
	\centering
	\includegraphics[width=0.92\linewidth]{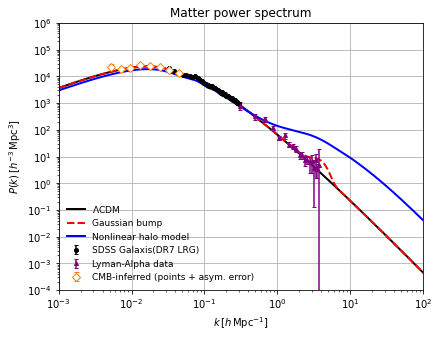}
	\caption{
	Linear matter power spectrum at $z=0$ for the standard
	$\Lambda$CDM model and the Gaussian-feature benchmark. The observational points are
	taken from the compilation of Chabanier et
	al.~\citep{2019MNRAS.489.2247C}, which combines constraints from the CMB, galaxy clustering, weak lensing, and the Ly$\alpha$ forest. The	benchmark feature approaches the small-scale region probed by the	Ly$\alpha$ data and is therefore treated as a boundary-motivated choice.}
	\label{fig:power_spectrum_constraints}
\end{figure}

Fig.~\ref{fig:power_spectrum_constraints} compares the standard and
modified linear matter spectra with the observational compilation. Our
parameters are chosen to fit the UVLF data:
\(
A_b=4.5 \,\mathrm{hMpc}^{-1}\,
\sigma_b=0.8 \,\mathrm{hMpc}^{-1}\,
k_\ast=4.15\,\mathrm{hMpc}^{-1}\)
place the enhancement close to the range probed by the Ly$\alpha$
forest without producing an obvious disagreement at the level of this
comparison. We therefore adopt this model as a boundary-motivated
benchmark that moves the halo response toward relatively large masses
while remaining qualitatively consistent with the compiled
matter-power measurements.


A critical consistency test for any cosmological scenario that enhances the abundance of early collapsed structures is its impact on the Epoch of Reionization (EoR). Because the localized Gaussian bump accelerates structure formation and increases the number density of halos hosting star-forming galaxies at $z \gtrsim 10$, it naturally elevates the early production rate of hydrogen-ionizing photons. The cumulative ionization history is tightly constrained by the Thomson scattering optical depth to the CMB measured by \textit{Planck}, $\tau_{\mathrm{reio}} = 0.054 \pm 0.007$ \citep{2020A&A...641A...6P}, which implies a midpoint of reionization at $z_{\mathrm{reio}} \sim 7.7$. While a broad, large-amplitude power enhancement across all small scales risks causing premature or ``over-reionization'' at $z > 10$, our localized feature is narrowly confined to $k_* \approx 4.15\,h\,\mathrm{Mpc}^{-1}$ and primarily enhances rare, high-mass halos ($M \gtrsim 10^{10}\,h^{-1}M_\odot$) rather than the faint, low-mass dwarf halo population ($M \lesssim 10^8\,h^{-1}M_\odot$) that dominates the integrated cosmic ionizing photon budget. Consequently, the moderate enhancement in the bright-end UVLF at $z \ge 11$ produces only a modest, perturbative boost to the early electron scattering optical depth ($\Delta \tau \lesssim 0.005$), fully compatible with the $1\sigma$ \textit{Planck} credible interval. Furthermore, any residual excess in ionizing photon production at high redshifts can be readily accommodated within standard astrophysical uncertainties by a mild redshift evolution in the ionizing photon escape fraction or the clumping factor of the intergalactic medium, without compromising the model's ability to fit the bright-end JWST observations.

The other issue is the safety from Primordial Black Holes (PBH) and spectral distortions.
We remark that the amplitude of the localized curvature power enhancement required here, 
$\mathcal{P}_\mathcal{R}(k_*) \sim \mathcal{O}(10^{-8}\text{--}10^{-7})$, is several orders of 
magnitude below the critical threshold for PBH formation, which 
typically demands $\mathcal{P}_\mathcal{R} \sim 10^{-2}\text{--}10^{-1}$ on small scales 
\citep{2020JCAP...04..007M,2025JCAP...09..045P}. Furthermore, since the feature is centered at $k_* \approx 4.15\,h\,\mathrm{Mpc}^{-1}$ 
(well outside the primary $\mu$-distortion window $k \sim 50\text{--}10^4\,\mathrm{Mpc}^{-1}$ which 
integrates power at much smaller acoustic damping scales), the resulting chemical potential 
$\mu \lesssim 10^{-8}$ and Compton $y$-distortion are entirely negligible and safely beneath 
the bounds set by {COBE}/FIRAS as well as future targets like PIXIE \citep{2017MNRAS.471.1126A,2024arXiv240612985C}.

\section{From Halo Abundance to the UV Luminosity Function}
\label{sec:uvlf_model}

The rest-frame ultraviolet luminosity function,
\(\Phi_{\rm UV}(M_{\rm UV},z)\), gives the comoving number density of
galaxies per unit UV absolute magnitude. We obtain it from the halo
mass function using the semi-analytical prescription of
references~\cite{2024JCAP...07..078C,2026MNRAS.546ag269C}. Only the ingredients
needed for the present calculation are summarized here.

For a deterministic halo-to-luminosity relation, the contribution from
galaxies in environment \(i\) is

\begin{equation}
	\Phi_{\rm UV}^{i}(M_{\rm UV},z)
	=
	\frac{1}{M_h}
	\frac{dn}{d\ln M_h}
	\left|
	\frac{dM_h}{dL_{\rm UV}^{i}}
	\right|
	\left|
	\frac{dL_{\rm UV}^{i}}{dM_{\rm UV}}
	\right|,
	\qquad
	i\in\{\mathrm{fb},\mathrm{nofb}\}.
	\label{eq:uvlf_mapping}
\end{equation}
where we have to types of environments, one with radiative feedback (fb) and one with no feedback (nofb) at all. The cosmological model enters through the halo mass function and the
remaining factors describe how halo mass is converted into UV
luminosity.

For a halo unaffected by photoionization feedback, the luminosity is
written as

\begin{equation}
	L_{\rm UV}^{\rm nofb}(M_h,z)
	=
	\frac{\epsilon_{\ast,10,\rm UV}}
	{K_{\rm UV,fid}}
	H(z)
	\left(\frac{\Omega_b}{\Omega_m}\right)
	\left(
	\frac{M_h}{10^{10}M_\odot}
	\right)^{\alpha_\ast}
	M_h ,
	\label{eq:uv_luminosity_nofb}
\end{equation}

where \(\epsilon_{\ast,10,\rm UV}\) fixes the normalization at
\(M_h=10^{10}M_\odot\), and \(\alpha_\ast\) controls its mass
dependence, $\Omega_b$ and $\Omega_m$ is the density parameter of baryon and matter respectively. We use

\[
K_{\rm UV,fid}
=
1.15485\times10^{-28}
\,
\frac{M_\odot\,{\rm yr}^{-1}}
{{\rm erg}\,{\rm s}^{-1}\,{\rm Hz}^{-1}} .
\]

In ionized regions, the available gas and hence the UV luminosity are
suppressed according to

\begin{equation}
	f_{\rm gas}(M_h)
	=
	2^{-M_{\rm crit}/M_h},
	\qquad
	L_{\rm UV}^{\rm fb}
	=
	f_{\rm gas}(M_h)L_{\rm UV}^{\rm nofb}.
	\label{eq:uvlf_feedback}
\end{equation}

The parameter \(M_{\rm crit}\) sets the halo-mass scale below which
photoionization feedback becomes important. The escape fraction $f_{\rm esc}$ of
ionizing photons is parameterized as

\begin{equation}
	f_{\rm esc}(M_h)
	=
	\epsilon_{\rm esc,10}
	\left(
	\frac{M_h}{10^{10}M_\odot}
	\right)^{\alpha_{\rm esc}} .
	\label{eq:escape_fraction}
\end{equation}

The luminosity is converted to an AB absolute magnitude through

\begin{equation}
	M_{\rm UV}
	=
	51.6
	-
	2.5\log_{10}
	\left(
	\frac{L_{\rm UV}}
	{{\rm erg}\,{\rm s}^{-1}\,{\rm Hz}^{-1}}
	\right).
	\label{eq:uv_ab_magnitude}
\end{equation}

Finally, galaxies in ionized and neutral regions are combined using

\begin{equation}
	\Phi_{\rm UV}
	=
	Q_{\rm HII}(z)\Phi_{\rm UV}^{\rm fb}
	+
	\left[1-Q_{\rm HII}(z)\right]
	\Phi_{\rm UV}^{\rm nofb},
	\label{eq:total_uvlf}
\end{equation}

where \(Q_{\rm HII}(z)\) is the ionized volume fraction. The full
calculation of the reionization history and its coupling to the galaxy
population is described in \cite{2026MNRAS.546ag269C}.


Our aim is to test whether the modified halo abundance alone can
increase the number of UV-bright galaxies. We therefore use the
low-redshift branch of the reference astrophysical model and keep it
fixed in both cosmologies:

\begin{equation}
	\boldsymbol{\theta}_{\rm astro}^{\rm std}
	=
	\boldsymbol{\theta}_{\rm astro}^{\rm G}
	=
	\boldsymbol{\theta}_{\rm astro}^{\rm low\text{-}z},
	\label{eq:fixed_astrophysical_parameters}
\end{equation}
where $	\boldsymbol{\theta}$ represent all astrophysical parameters.
The parameters are not refitted to the high-redshift JWST data. The
only quantity changed between the two predictions is the halo mass
function. A similar idea is used while studying luminosity function of galaxies in voids \citep{2025ApJ...988..271A}.
\begin{table}
\caption{
		Fixed astrophysical parameters used in the halo-to-UV mapping \citep{2026MNRAS.546ag269C}.
		Logarithms are base ten.
	}
	\label{tab:fixed_uvlf_parameters}
	\centering
	\setlength{\tabcolsep}{3pt}
		\begin{tabular}{
				p{0.30\columnwidth}
				p{0.20\columnwidth}
				p{0.38\columnwidth}
		}
			Parameter
			& Adopted value
			& Role
		\\
		\hline
			\(\log_{10}
			\epsilon_{\ast,10,\rm UV}\)
			& \(-0.889\)
			& UV-luminosity normalization
			\\
			\(\alpha_\ast\)
			& \(0.302\)
			& Mass dependence of the effective stellar efficiency
			\\
			\(\log_{10}
			\epsilon_{\rm esc,10}\)
			& \(-0.810\)
			& Escape-fraction normalization
			\\
			\(\alpha_{\rm esc}\)
			& \(-0.17\)
			& Mass dependence of the escape fraction
			\\
			\(\log_{10}
			(M_{\rm crit}/M_\odot)\)
			& \(10.11\)
			& Photoionization-feedback mass scale
		\end{tabular}
\end{table}

The reference model permits the UV-efficiency parameters to change
between low and high redshift branches. Here we instead impose

\begin{equation}
	(\log_{10}\epsilon_{\ast,10,\rm UV})_{\rm hi}
	=
	(\log_{10}\epsilon_{\ast,10,\rm UV})_{\epsilon,\rm lo}
	=
	-0.910,
	\qquad
	\alpha_{\rm hi}
	=
	\alpha_{\rm lo}
	=
	0.303.
	\label{eq:no_astrophysical_evolution}
\end{equation}

There is no additional high-redshift increase in the UV efficiency.

\label{sec:uvlf_results}

\begin{figure*}
\centering
	\includegraphics[
	width=1.5\textwidth,
	height=0.9\textheight,
	keepaspectratio
	]{\detokenize{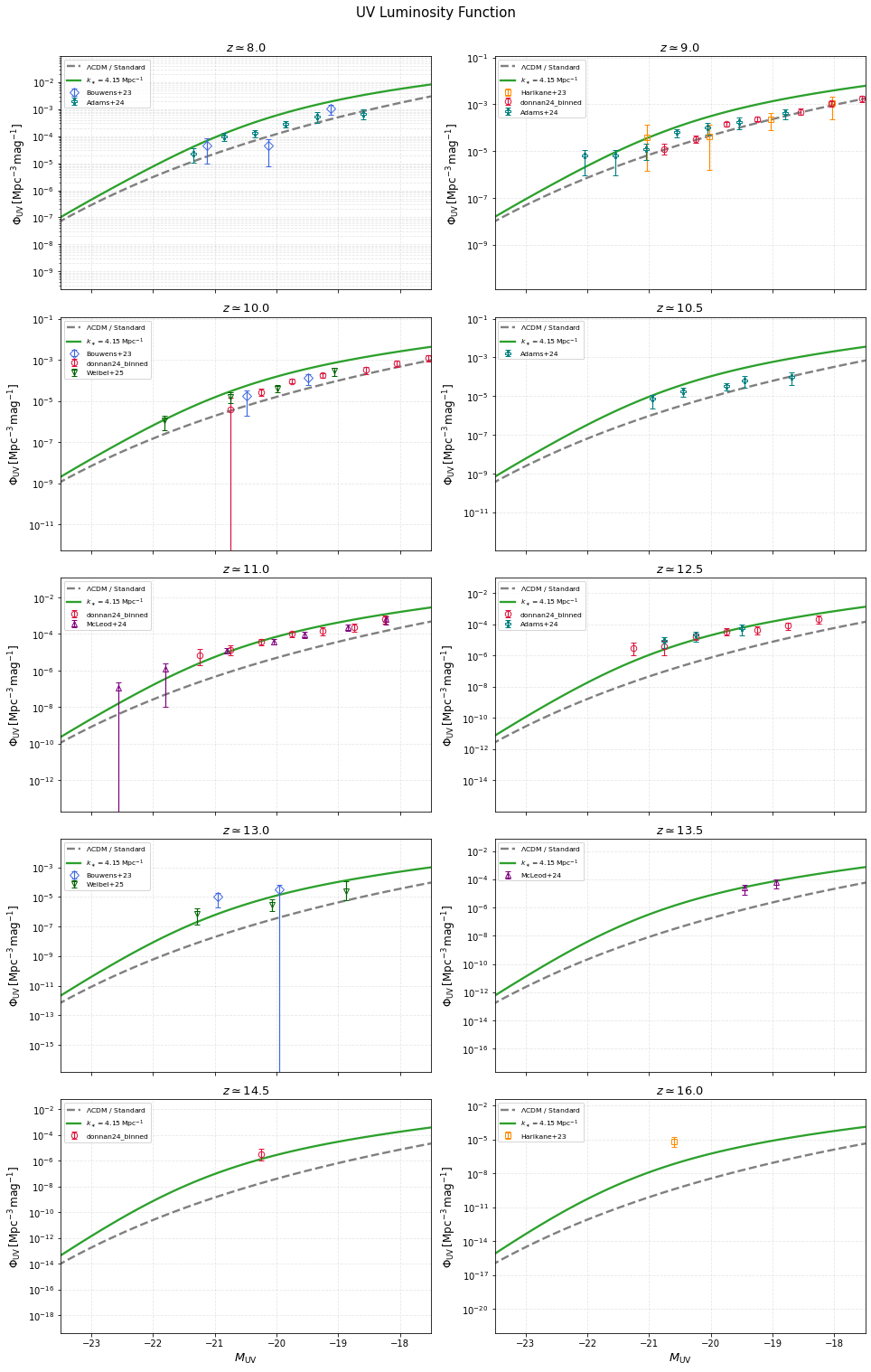}}
	\caption{
		Comparison of the ultraviolet luminosity function in the
		Gaussian-feature model and the standard \(\Lambda\)CDM model at
		several redshifts. The solid black curves show the modified model,
		while the dashed gray curves show the standard prediction. The
   	colored points are observational measurements from
	~\citep{2023MNRAS.523.1009B,2024ApJ...965..169A,2023MNRAS.518.6011D,2024MNRAS.533.3222D,2024MNRAS.527.5004M,2025ApJ...980..138H,2026ApJ..1002..136W}.
	}
	\label{fig:uvlf_number_density_galaxies}
\end{figure*}

Fig.~\ref{fig:uvlf_number_density_galaxies} compares the standard
and Gaussian-feature predictions with high-redshift UVLF measurements.
Because the same astrophysical parameters are used for both curves,
their separation directly reflects the change in halo abundance.
The relative enhancement also increases with redshift. Defining
\(\mathcal{R}_{\rm UVLF}
=\Phi_{\rm UV}^{\rm G}/\Phi_{\rm UV}^{\rm std}\), we find that the
Gaussian-feature model produces a progressively larger excess over the
standard prediction at earlier times. This trend follows the behavior
of the halo mass function, and at higher redshift, the
halos associated with UV-bright galaxies lie deeper in the rare-object
tail, so a fixed increase in the mass variance produces a larger
fractional change in their abundance. 

The Gaussian feature generally raises the predicted UVLF, with the
largest effect appearing at higher redshift and toward the bright end.
At \(z\simeq8\)--\(10\), the standard prediction remains competitive,
while from \(z\simeq11\) onward the Gaussian model is typically closer
to several of the observed points. The modification does not reproduce
every measurement, but it moves the prediction in the direction needed
to reduce the bright-galaxy deficit without retuning the astrophysical
model.

The comparison at \(z=14.5\) and \(z=16\) should be treated cautiously,
since only one observational point is available at each redshift. These
points indicate the direction of the model response but do not provide
an independent statistical test.

\label{subsec:uvlf_chi_square_comparison}

To supplement the visual comparison, we evaluate a simple statistic in
logarithmic UVLF space:

\begin{equation}
	\chi^2_{\log}
	=
	\sum_i
	\left[
	\frac{
		\log_{10}\Phi_{\rm model}(M_{{\rm UV},i},z)
		-
		\log_{10}\Phi_{\rm obs}(M_{{\rm UV},i},z)
	}{
		\sigma_{\log,i}
	}
	\right]^2 .
	\label{eq:uvlf_log_chi_square}
\end{equation}

The theoretical curve is interpolated to the magnitude of each
observational point. For asymmetric measurements, the upper or lower
uncertainty is selected according to the direction of the model
residual. This statistic is used only to compare the fixed curves; it
is not a full likelihood analysis.
\begin{table*}
	\centering
	\caption{
		Quantitative comparison of the best model at each redshift using
		\(\chi^2_{\log}\). Only redshifts with at least two observational
		points are included.
	}	\label{tab:uvlf_chi_square_best}
		\begin{tabular}{ccccc}			\(z\)			& Best model
			& \(N\)
			& \(\chi^2_{\log}/N\)
			\\
			\hline
			\(8.0\)
			& \(\Lambda{\rm CDM}\)
			& \(9\)
			& \(2.9\)
			\\
			\(9.0\)
			& \(\Lambda{\rm CDM}\)
			& \(18\)
			& \(1.4\)
			\\
			\(10.0\)
			& \(\Lambda{\rm CDM}\)
			& \(13\)
			& \(5.4\)
			\\
			\(10.5\)
			& \(\Lambda{\rm CDM}\)
			& \(5\)
			& \(2.6\)
		\\
			\(11\)
			& Gaussian 
			& \(14\)
			& \(1.9\)
			\\
			\(12.5\)
			& Gaussian 
			& \(10\)
			& \(1.4\)
			\\
			\(13.0\)
			& Gaussian 
			& \(5\)
			& \(1.6\)
			\\
			\(13.5\)
			& Gaussian 
			& \(2\)
			& \(0.1\)
		\end{tabular}
\end{table*}

The preference for the standard model at \(z\simeq8\)--\(10\) is not
unexpected. The astrophysical parameters used here were taken from the
low-redshift branch of the reference model and were kept fixed when the
Gaussian feature was introduced. They are therefore naturally better
suited to the regime in which the standard \(\Lambda\)CDM model was
originally calibrated. 
A more complete analysis would vary the cosmological and astrophysical
parameters simultaneously. This would require a joint likelihood or
Monte Carlo Markov Chain (MCMC) analysis over the Gaussian-feature parameters
\((A_b,\sigma_b,k_\ast)\), together with
\(\epsilon_{\ast,10,\rm UV}\), \(\alpha_\ast\),
\(\epsilon_{\rm esc,10}\), \(\alpha_{\rm esc}\), and
\(M_{\rm crit}\). Such an analysis could identify a single parameter
region that balances the lower-redshift UVLF constraints with the
enhanced abundance required at higher redshift. It may therefore
improve the agreement of the Gaussian model at \(z\simeq8\)--\(10\),
although this cannot be established without performing the joint fit.
The present calculation should instead be viewed as a controlled test
of the cosmological contribution, with the astrophysical mapping held
fixed.

%
%
%
%
%
%
%
%

\section{Conclusions and Future Prospects}
\label{sec:conclusions}

In this work, we investigated whether a localized enhancement in the primordial curvature power spectrum can alleviate the tension between standard $\Lambda\mathrm{CDM}$ cosmological predictions and recent JWST observations of an overabundance of luminous, massive galaxies at $z \gtrsim 10$. Employing the framework of EST with the Sheth--Tormen ellipsoidal collapse barrier, we introduced a Gaussian bump characterized by amplitude $A_b$, scale $k_*$, and width $\sigma_b$, motivated by transient departures from slow-roll inflation or resonant particle production in the early Universe.
Our main findings are as follows.
The impact of a localized feature in the primordial power spectrum on the dark matter halo mass function (HMF) is non-trivially amplified toward higher redshifts. Because the critical collapse threshold $\delta_c(z) \propto D(z)^{-1}$ increases monotonically with $z$, the abundance of rare, extreme-mass halos at the exponentially suppressed high-$\nu$ tail becomes exceptionally sensitive to fractional changes in the smoothed mass variance, $S(M)$. Consequently, a moderate boost in power around $k_* \sim 2\text{--}4\,h\,\mathrm{Mpc}^{-1}$ generates a large relative enhancement ($R_{\mathrm{HMF}} \gg 1$) at $z \ge 11$, while remaining well within small-scale constraints at lower redshifts.
EST based progenitor distributions demonstrate that the presence of the Gaussian bump accelerates the assembly history of high-redshift halos, leading to systematically higher median half-mass formation redshifts ($\Delta z_{\mathrm{half}} > 0$). This provides a natural physical environment for early, concentrated star formation.

Observational constraints from the 1D Ly$\alpha$ flux power spectrum at $z \sim 2\text{--}4$ tightly delimit the permissible parameter space. Our comparison with the reconstructed linear matter power spectrum indicates that substantially larger amplitudes would become increasingly difficult to reconcile with the small-scale power constraints, motivating \(A_b=8\,h\,{\rm Mpc}^{-1}\) as a boundary-motivated benchmark rather than as a formal upper limit.

 Our benchmark parameters  ($A_b \le 8\,h\,\mathrm{Mpc}^{-1}$ for $k_* \approx 4.15\,h\,\mathrm{Mpc}^{-1}$) respect these bounds while maximizing the allowed boost at the scale corresponding to the collapse of halos hosting UV-bright galaxies at $z \gtrsim 10$.

Propagating the modified HMF through a semi-analytic halo-to-luminosity relation with fixed standard astrophysical parameters, we find a substantial improvement in the goodness-of-fit to the observed JWST UV luminosity functions. As shown by our $\chi^2$ analysis, while standard $\Lambda\mathrm{CDM}$ provides a competitive fit at $z \le 10$, the Gaussian feature model decisively outperforms $\Lambda\mathrm{CDM}$ at $z = 11\text{--}13.5$, naturally shifting the bright-end cutoff toward observed densities without invoking extreme or unphysical star formation efficiencies ($f_* > 1$).
We emphasize that localized primordial power enhancement should be viewed as a \textit{partial, complementary cosmological contribution} rather than a standalone silver bullet. While the model significantly bridges the gap at the highest redshifts, fully reconciling the observations likely requires a synergistic combination of:
Primordial/cosmological modifications that boost the baseline abundance of massive potential wells; and
non-standard astrophysical mechanisms operating at low metallicity and high gas densities, such as feedback-free starbursts, top-heavy stellar initial mass functions (IMFs), or non-negligible contributions from obscured AGN.\\
Several prospective avenues will test and refine the scenario explored in this work:\\
Upcoming large-area JWST spectroscopic surveys (e.g., via NIRSpec PRISM/grism programs) and the Nancy Grace Roman Space Telescope High Latitude Wide Area Survey will firmly establish spectroscopic redshifts, eliminate low-$z$ interloper contamination, and constrain cosmic variance at $z > 10$.\\
An enhanced population of early massive halos necessarily alters the production rate of ionizing photons during the Cosmic Dawn. Forthcoming 21-cm power spectrum measurements from the Square Kilometre Array (SKA) and HERA, alongside precision optical depth measurements ($\tau_{\mathrm{reio}}$) from future CMB missions (e.g., LiteBIRD), will place orthogonal constraints on the earliest phases of structure formation and rule out over-reionization scenarios.

Sharp features in the primordial power spectrum often induce characteristic scale-dependent non-Gaussianities ($f_{\mathrm{NL}}^{\mathrm{local}}$ or oscillatory bispectra). Detailed consistency checks combining excursion set peak statistics with non-Gaussian initial conditions will provide an explicit consistency relation between the JWST bright-end excess and large-scale structure 3-point correlation functions.\\
Direct cosmological zoom-in and volume simulations initialized with localized power bumps will allow a self-consistent treatment of baryonic feedback, gas accretion rates, and star formation quenching at $z > 10$, going beyond semi-analytic approximations and shedding light on the morphology and clustering of these first-generation galaxies.

\section*{Data availability}
This study does not involve the use or production of original data.

\section*{Acknowledgments}
The authors used ChatGPT (OpenAI) for assistance with language editing and manuscript presentation.
SB is partially supported by the Abdus Salam International Center for Theoretical Physics (ICTP) under the regular associateship scheme.
Moreover, SB is partially supported by the Sharif University of Technology Office of Vice President for Research under Grant No. G4010204.




\bsp	
\label{lastpage}
\end{document}